\pdfoutput=1

\documentclass[twocolumn,trackchanges]{aastex62}
\graphicspath{{./figures/}}

\usepackage{multirow}
\usepackage{booktabs}
\usepackage{amstext}

\usepackage{amsmath}
\newcommand{\ie}{{\it i.e.}}

\newcommand{\eg}{{\it e.g.}}

\newcommand{\cf}{{\it cf.}}

\newcommand{\Tab}{Tab.}

\newcommand{\TXS}{TXS 0506+056}

\newcommand{\figu}[1]{Fig.~\ref{fig:#1}}
\newcommand{\tabl}[1]{\Tab~\ref{tab:#1}}

\definecolor{deepmagenta}{rgb}{0.8, 0.0, 0.8}
\newcommand{\revise}[1]{#1}

\shorttitle{Models for the historical flare of TXS 0506+056}
\shortauthors{X. Rodrigues et al.}

\begin{document}

\title{Leptohadronic blazar models applied to the 2014-15 flare of TXS 0506+056}

\correspondingauthor{Xavier Rodrigues, Shan Gao}
\email{xavier.rodrigues@desy.de, shan.gao@desy.de}

\author[0000-0001-9001-3937]{Xavier Rodrigues}
\affiliation{Deutsches Elektronen-Synchrotron (DESY), Platanenallee 6, 15738 Zeuthen, Germany}
\nocollaboration

\author[0000-0002-5309-2194]{Shan Gao}
\affiliation{Deutsches Elektronen-Synchrotron (DESY), Platanenallee 6, 15738 Zeuthen, Germany}
\nocollaboration

\author[0000-0003-2837-3477]{Anatoli Fedynitch}
\affiliation{Deutsches Elektronen-Synchrotron (DESY), Platanenallee 6, 15738 Zeuthen, Germany}
\nocollaboration

\author[0000-0001-7203-5366]{Andrea Palladino}
\affiliation{Deutsches Elektronen-Synchrotron (DESY), Platanenallee 6, 15738 Zeuthen, Germany}
\nocollaboration

\author[0000-0001-7062-0289]{Walter Winter}
\affiliation{Deutsches Elektronen-Synchrotron (DESY), Platanenallee 6, 15738 Zeuthen, Germany}
\nocollaboration

\begin{abstract}
We investigate whether the emission of neutrinos observed in 2014-15 from the direction of the blazar \TXS{} can be accommodated with leptohadronic multi-wavelength models of the source commonly adopted for the 2017 flare. While multi-wavelength data during the neutrino flare are sparse, the large number of neutrino events ($13\pm5$) challenges the missing activity in gamma rays. We illustrate that {two to five} neutrino events during the flare can be explained with leptohadronic models of different categories: a one-zone model, a compact core model, and an external radiation field model. If, however, significantly more events were to be accommodated, the predicted multi-wavelength emission levels would be in conflict with observational X-ray constraints, or with \revise{the high-energy gamma ray fluxes observed by the Fermi LAT}, depending on the model. For example, while the external radiation field model can predict up to five neutrino events without violating X-ray constraints, the absorption of \revise{high-energy} gamma rays is in {minor} tension with data. We therefore do not find any model that can simultaneously explain the high event number quoted by IceCube and the (sparse) electromagnetic data during the neutrino flare.
\end{abstract}

\keywords{Multi-messenger, blazar, astrophysical neutrinos, TXS 0506+056}


\section{Introduction}

\label{sec:intro}

The object \TXS{} is an Active Galactic Nucleus (AGN) of the blazar type, classified as a BL Lac {object}, with a measured redshift of $z=0.3365$~\citep{2018arXiv180201939P}. 
In September 2017, a muon neutrino with a reconstructed energy of about 290~TeV was observed by IceCube from a position compatible with this source in coincidence with a period of flaring in multiple wavelengths~\citep{TXS_MM} {at a significance level of $3\sigma$}. This event has enticed the multi-messenger community to explore the potential of {\TXS{}} as a source of astrophysical neutrinos.
The connection between neutrino production and the electromagnetic flare has been described by several leptohadronic ($p\gamma$) production models~\citep{Gao:2018mnu,Cerruti:2018tmc,Zhang:2018xrr,Keivani:2018rnh,Ahnen:2018mvi,Sahakyan:2018voh,Gokus:2018lgx,Righi:2018xjr}, as well as hadronic ($pp$) production models~\citep{Liu:2018utd,Sahakyan:2018voh}. For example, it has been concluded by \citet{Gao:2018mnu} that conventional one-zone models describing the spectral energy distribution and the neutrino event suffer from too low neutrino rates in combination with excessively high neutrino energies or  sustained super-Eddington injection luminosities. The current theoretical consensus is that the geometry of the radiation zone must be more complex, involving a compact radiation core with high photohadronic interaction rates \citep{Gao:2018mnu}, or external radiation fields boosted into the jet frame, either thermal \citep{Keivani:2018rnh} or non-thermal \citep{Ahnen:2018mvi}.

Triggered by the multi-messenger discovery of the 2017 flare, IceCube searched their archival data for an excess from the direction of {\TXS{}}~\citep{TXS_orphanflare} for the entire duration of IceCube's data taking. In the period between October 2014 and March 2015, a temporal clustering has been detected of 64 events in total within $3^\circ$ of the direction of the {same source}. By using a likelihood function, in which the atmospheric background is taken into account and the signal is assumed to be distributed as a power law, a 3.5$\sigma$ excess over the atmospheric background was found, with an estimated number of signal events of $13\pm5$ (henceforth the ``historical neutrino flare''). The most energetic event has a deposited energy of 20 TeV in IceCube, while most events have energies around $\sim$ 10 TeV. Interestingly, this signal was not accompanied by any significant increase in electromagnetic emission. In contrast to the 2017 flare, the multi-wavelength data from this period are very sparse and the only constraints on the spectral energy distribution (SED) can be derived from gamma-ray flux measurements by the Fermi LAT~\citep{Aartsen:2019gxs}, as well as radio and optical monitoring data compiled by~\citet{Padovani:2018acg}. Additionally, the Swift Burst Alert Telescope (BAT) that monitors X-ray transients and performs regular sky surveys, was not  triggered during the period of the neutrino flare and did not detect {\TXS{}} in the 15--50 keV band, implying that its flux during the neutrino flare was significantly less than 3 mCrab, or $7.2 \times 10^{-11} \text{erg}~\text{cm}^{-2}~\text{s}^{-1}$~\citep{Krimm:2013lwa}. 
Based on \textit{Fermi} data, \citet{Padovani:2018acg} have speculated that there may be a hardening in the SED of the source above 2~GeV during the neutrino flare, although this feature may in fact not be significant~\citep{Aartsen:2019gxs}. Theoretical models for the historical flare are sparse, facing the challenge that the high neutrino flux has to be accommodated with the inconspicuous SED activity~\citep{Murase:2018iyl}. A possible way out could be 
jet-cloud/star ($pp$) {interactions~\citep{Bednarek:1996ffa,Barkov:2010km,Wang:2018zln}}, whereas the photohadronic model in~\citet{Halzen:2018iak} does not contain a self-consistent SED computation. 

In this letter, we present a theoretical analysis of the neutrino and electromagnetic emission during the historical flare of {\TXS{}}. We focus on the available observational evidence during the neutrino emission period only -- which was presented above. Due to the lack of enhanced gamma-ray activity, we treat the historical flare independently from the 2017 event. Motivated by the limited constraints from long-term multi-wavelength data, we do not attempt to derive a time-dependent model that explains the transitions between the neutrino bright and quiet states. The multi-messenger SED is computed using the self-consistent numerical code {\sc AM}$^3$~\citep{2017ApJ...843..109G}, that has been successfully applied in the interpretation of the 2017 flare \citep{Gao:2018mnu}, and that has been extended by the inclusion of external radiation fields. Apart from a conventional one-zone model, we test two other classes, namely an inverse-Compton dominated compact-core model and a model involving an external radiation field from accretion disk radiation isotropized in a broad-line region (BLR)-- {considering a scenario in which this source} possesses features typical of Flat-Spectrum Radio Quasars (FSRQs), that are out-shined by the radiation from the jet.

\section{Methods}

We construct the models for the simultaneous emission of neutrinos and photons using the leptohadronic code {\sc AM}$^3$, solving self-consistently the time-dependent kinetic equations for non-thermal electrons, positrons, protons, neutrons, photons and neutrinos produced in the relativistic jet. The production region is simulated as a spherical blob of radius $R'_\mathrm{blob}$ in its rest frame\footnote{Primed quantities refer to the blob rest frame, unprimed quantities to the observer's frame}, moving along the blazar jet with a bulk Lorentz factor $\Gamma_\mathrm{b}$. The assumption is that protons and electrons are accelerated to a power-law spectrum $dN/dE^\prime\propto E^{\prime-2}$, up to certain maximum Lorentz factors $\gamma_{e,\mathrm{max}}^{\prime}$ and $\gamma_{p,\mathrm{max}}^{\prime}$, and are then injected isotropically into a radiation zone of the jet. They interact with the target photons according to \citet{Hummer:2010vx}, producing charged and neutral pions that ultimately decay into neutrinos and secondary gamma rays, electrons and positrons. These particles will feed into the electromagnetic cascade and potentially lead to signatures in the SED. The other interactions included in the model are electron synchrotron emission and synchrotron self-absorption, inverse Compton (IC) scattering by both electrons and protons, photon pair production and annihilation, and Bethe-Heitler pair production, $p\,\gamma\,\rightarrow\, p\,e^+\,e^-$. The electron synchrotron emission depends on the strength of the turbulent magnetic field in the radiation zone, $B'$, considered randomly oriented.

The critical quantity of interest is the number of neutrinos { predicted by the model}, which we compute by folding the {emitted} fluence with the effective area given \revise{by \citet{TXS_orphanflare} for the IC86b data period}. For the single neutrino observed during the 2017 flare, the expected number of {detectable} neutrinos is {likely} smaller than one for different reasons (Eddington bias in~\citet{Strotjohann:2018ufz} or too many associations expected in~\citet{Palladino:2018lov}). These arguments do not apply to the historical flare, where the predicted event number needs to be significantly larger than one to be compatible with observations. Since for blazars the number density of target photons (X-rays for {this particular source}) is low compared to other compact objects such as gamma-ray bursts, the optical depth to $p\gamma$ interactions is typically much lower than unity -- which needs to be compensated by a large proton {loading} in order to become a significant neutrino source. The photohadronic interaction rate can be enhanced by assuming a smaller production region or a higher density in X-rays. The latter can be achieved by an external radiation field boosted into the blob frame, such as the X-ray photons initially produced by the accretion disk and scattered by the dust or cloud surrounding the jet.

We perform extensive scans within physically plausible ranges for the parameter space {(for $10^{15.0}<R^{\prime}_\mathrm{blob}/\mathrm{cm}<10^{17.0}$, $10^{-3}<B^{\prime}/G<10$, $5<\Gamma_\mathrm{b}<50$, $0.06<E_\mathrm{p,inj}^{\prime}/\mathrm{PeV}<15$ and $10^{2.8}<L_\mathrm{p,inj}/L_\mathrm{e,inj}<10^{6.3}$)}. Yet, we cannot claim completeness of our scans because of the complexity of the problem. {The parameter space was searched using two methods: a grid-based parameter scan and a genetic algorithm \citep{goldberg89}. The {goodness of fit for each parameter set is defined according to a simple $\chi^2$-criterion in $\nu F_\nu$ between the simulated SED, and the optical, gamma-ray and X-ray constraints}}. \revise{Since a rigorous minimization is not feasible due to the sparsity of the data, we choose the ``neutrino-loudest'' areas of the parameter space.}\footnote{{A lepto-hadronic blazar model involves $N\sim10$ parameters. {While some parameters are correlated, a $\chi^{2}$ goodness of fit estimator produces a highly degenerate likelihood space given the underconstrained nature of the problem (only seven data points).}}} The emitted neutrino spectra, which significantly differ from power-laws, are convolved with the effective area of IceCube at the declination of the source~\citep[IC86b data period,][]{TXS_orphanflare} to obtain the predicted number of muon track events, which is then compared to the observed signal.

\section{Results}

\begin{table*}[btp]
\caption{Selected parameter sets for each model, predicted number of neutrino events $N_\nu$ and SED quality compared to data. \revise{The corresponding neutrino rate is given by $N_{\nu}/T$, where $T=158~\mathrm{days}$ is the duration of the neutrino flare.} The values for 1-zone (a) and (b) are given for two representative curves from \figu{onezone} (red curve from the left panel and green curve from the right panel). The physical luminosities $L_{e, \mathrm{phys}}^\mathrm{obs}$ and $L_{p, \mathrm{phys}}^\mathrm{obs}$ carried by electrons and protons are given by $L_\mathrm{phys}^\mathrm{obs}=L_\mathrm{iso}/\Gamma^2$, where $L_\mathrm{iso}$ is the isotropic-equivalent luminosity. {The physical luminosities can be compared to the Eddington luminosity, which is $4\times10^{46}~\mathrm{erg/s}$ for a black hole mass of $3\times10^8~M_\odot$ estimated by \citet{Padovani:2019xcv}; note, however, that this value can be temporarily exceeded \revise{during} flares.}}
\label{tab:parameters}
\begin{center}
\begin{tabular}{lrlrrrrrrrrrr}
  \toprule
  &\multicolumn{2}{c}{Quality criteria}&\multicolumn{10}{c}{Parameters} \\
  \cmidrule{2-3} \cmidrule(ll){4-13}
   &  $N_\nu$ & SED & $B'$ & $R'_\mathrm{blob}$ & $\Gamma_\mathrm{b}$ & $\gamma^\mathrm{max}_\mathrm{p,\,  obs}$ & $L_{e,\mathrm{phys}}^\mathrm{obs}$ & $L_{p,\mathrm{phys}}^\mathrm{obs}$ & {$f_\mathrm{esc}$} & $T_\mathrm{disk}$ & $L_\mathrm{disk}$ & $R_\mathrm{BLR}$  \\
  Model & & & [G] & [cm] & & & {[$\mathrm{erg/s}$]} & {[$\mathrm{erg/s}$]} & & [K] & [$\mathrm{erg/s}$] &  [cm] \\
  \midrule
  1-zone (a) & $1.8$  & Compatible & $1.0$ & $10^{17}$ & $9.0$ & $10^{5.8}$ & {$10^{44.2}$} & {$10^{49.6}$} &{$10^{-2.5}$}&--&-- &-- \\
  {1-zone (b)} & {$13.2$} & {Overshoot}  & {$0.001$}  & {$10^{15}$} & {$7.0$} & {$10^{5.7}$} & {$10^{44.8}$} & {$10^{50.7}$} &{$10^{-2.5}$}&--&-- & -- \\
  \cmidrule{1-13}
  C. core (blob) & $0.0$               &\multirow{2}{*}{Compatible}& \multirow{2}{*}{$0.01$} & $10^{18.7}$ & \multirow{2}{*}{$10.0$} & --         & {$10^{44.8}$}      & \revise{--} & \revise{$10^{-2.5}$} &--&-- & -- \\
  C. core (core) & $1.9$               &                           &                         & $10^{15}$        &                         & $10^{6.1}$ & {$10^{43.7}$} & {$10^{49.5}$} &{$10^{-2.5}$} &--&-- & --\\
  \cmidrule{1-13}

  Ext. field (a) &{4.9}& {Compatible} &  {0.6} & $10^{15.8}$ & {49.1} & $10^{5.9}$ & {$10^{43.6}$} & {$10^{48.7}$} & {$10^{-4.8}$} &      {$10^{5.7}$} & {$10^{44.7}$} & {$10^{17.8}$}\\
  Ext. field (b) &{4.0}& {Cutoff, 10~GeV}     &  {0.9}& {$10^{16.3}$} &{48.0} & {$10^{6.3}$} & {$10^{42.9}$}& {$10^{48.4}$} & {$10^{-3.1}$} &      {$10^{5.3}$}& {$10^{44.7}$} & {$10^{17.3}$}\\


    \bottomrule
\end{tabular}
\end{center}
\end{table*}

\begin{figure*}[btp]
	\centering
	\includegraphics[width=\columnwidth]{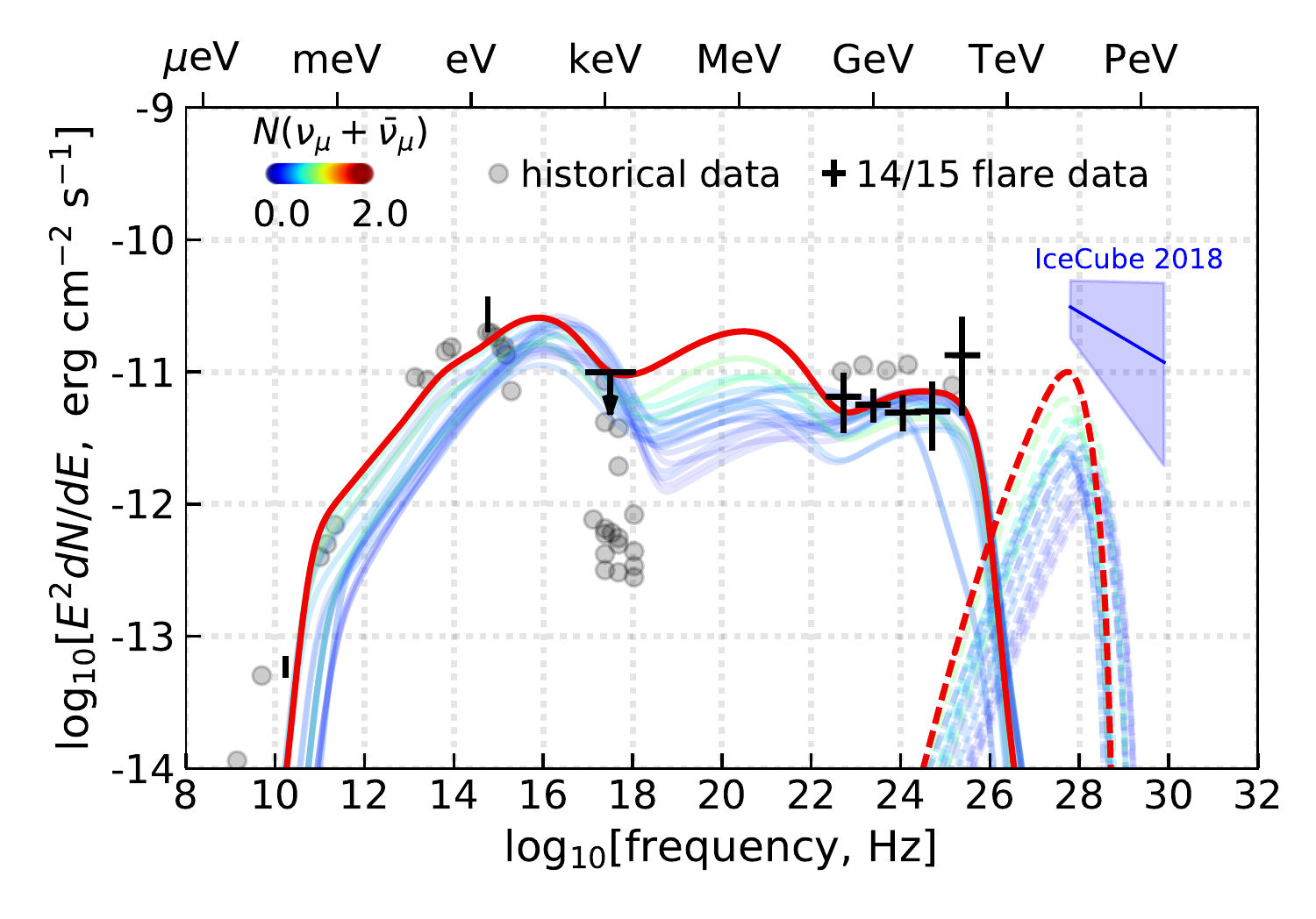}
	\includegraphics[width=\columnwidth]{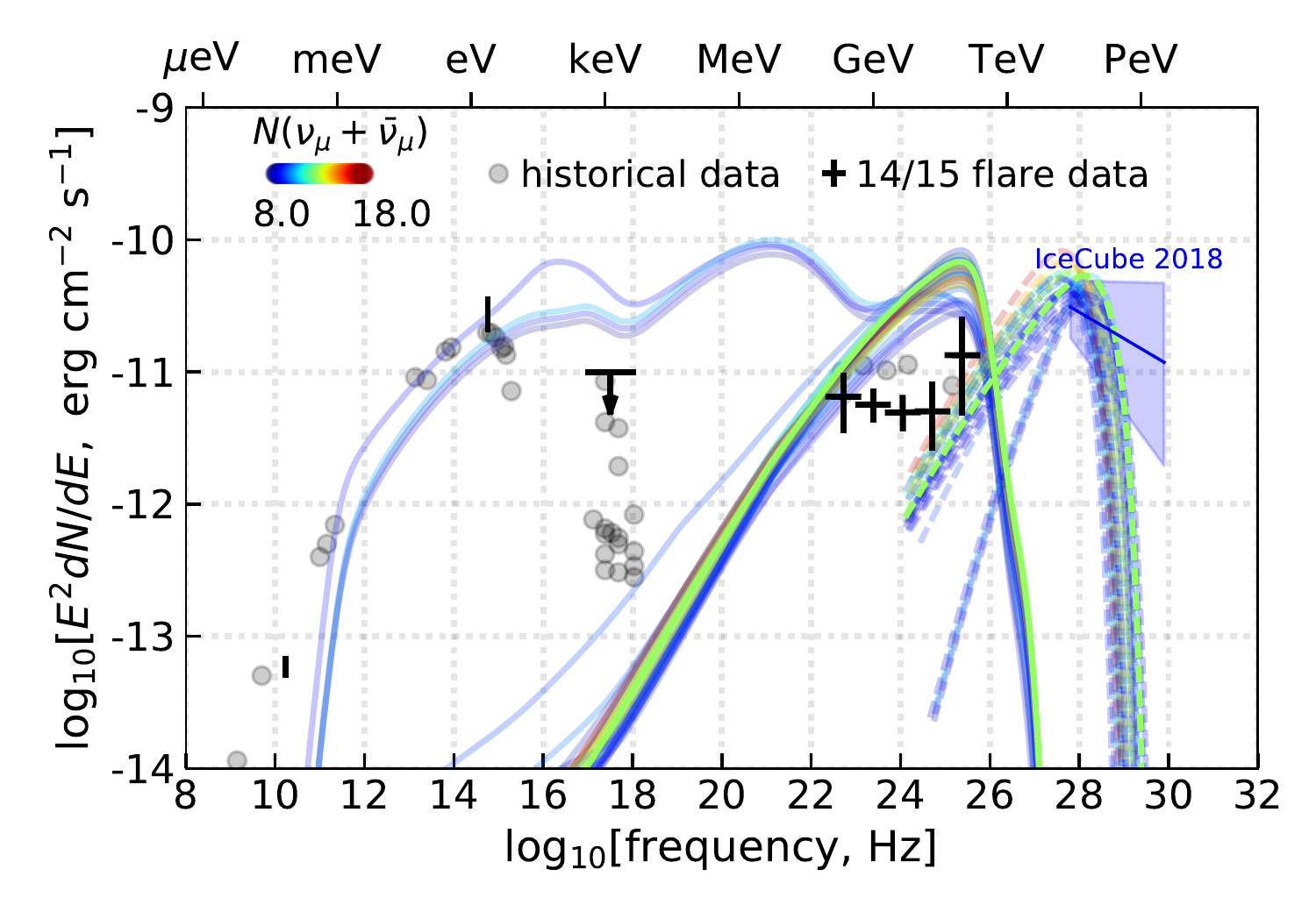}
	\caption{Spectral energy distributions (SEDs) and muon neutrino fluxes predicted by the one-zone hadronic model, compared to the single-flavor flux derived by IceCube~\protect{\citep{TXS_orphanflare}}. In the left panel, the parameter sets optimized to describe the SED in agreement with observations fail to explain neutrino emission; in the right panel the parameter sets account for $13\pm5$ {muon neutrinos in IceCube, but overshoot the multi-wavelength emission. \protect{\tabl{parameters}} contains the parameters for the red curve from the left panel, and the \revise{green} curve from the right panel. The observations available during the historical neutrino flare are plotted in black (see main text) and include one radio point \protect{\citep{Padovani:2018acg}}. The {archival} data taken during the years before 2017 from the databases of the Space Science Data Center (SSDC) and the NASA/IPAC Extragalactic Database (NED) are shown in gray. \label{fig:onezone}}}
\end{figure*}

We first test a conventional one-zone model, where the radiation zone consists of a single spherical blob. The neutrinos in the jet escape the blob over the free-streaming timescale $t'_\mathrm{FS}=R'_\mathrm{blob}/c$ (and likewise for the photons and neutrons that survive the interactions). Charged particles escape the blob at a slower rate due to the magnetic confinement. For simplicity, we implement an {energy-independent escape rate} for charged particles of $t'_\mathrm{esc}=f_\mathrm{esc}\,t'_\mathrm{FS}$, with \revise{$f_\mathrm{esc}>1$.}\footnote{{We also tested alternative scenarios; (a) $f_\mathrm{esc}=1$ and (b) a harder proton injection spectrum $dN/dE\propto E^{-1.5}$. Both {yield similar SEDs to that illustrated by the red curve in the left panel of \figu{onezone} if} $L_\mathrm{p,inj}$ is increased \revise{by a factor of 45} in case (a), and decreased \revise{by a factor of 3.8} in case (b).}} \figu{onezone} clearly demonstrates that the models compatible with the SED (left panel) produce too few neutrinos, where at most 1.8 events are expected during the duration of the neutrino flare (red curve, parameters listed in \tabl{parameters}). This number is limited by the X-ray constraint on the SED, which we derive from the non-detection by Swift BAT. The two bumps around the X-ray limit come from synchrotron and IC emission off e$^\pm$ that originate from $\gamma\gamma$ annihilation at higher energies, and from Bethe-Heitler pair production. This example demonstrates the importance of electromagnetic data across the entire spectrum to constrain theoretical models, since the electromagnetic cascade accompanying the neutrino production can be hidden in unconstrained energy ranges (such as MeV).

On the other hand, a compatible neutrino flux level implies an SED that is in tension with observations (right panel). These neutrino-compatible SEDs belong to a class of models with a strong hadronic cascade. Note that the self-consistently computed SED is very different from the \textit{ad hoc} assumption in \citet{Halzen:2018iak}, and peaks at lower energies. We also find a cluster of strongly IC-dominated solutions, due to a compact emission region and low magnetic field strength, supporting a high $p\gamma$ efficiency and hence higher neutrino fluxes. These solutions cannot, however, explain the emission outside the Fermi LAT range. On the other hand, the subset of models with sufficient synchrotron emission fail to simultaneously comply with the X-ray and gamma-ray constraints. We conclude that one-zone models are in tension with observations of the historical flare. The corresponding model parameters are listed in \tabl{parameters}.

\begin{figure}[btp]
	\centering
	\includegraphics[width=\columnwidth]{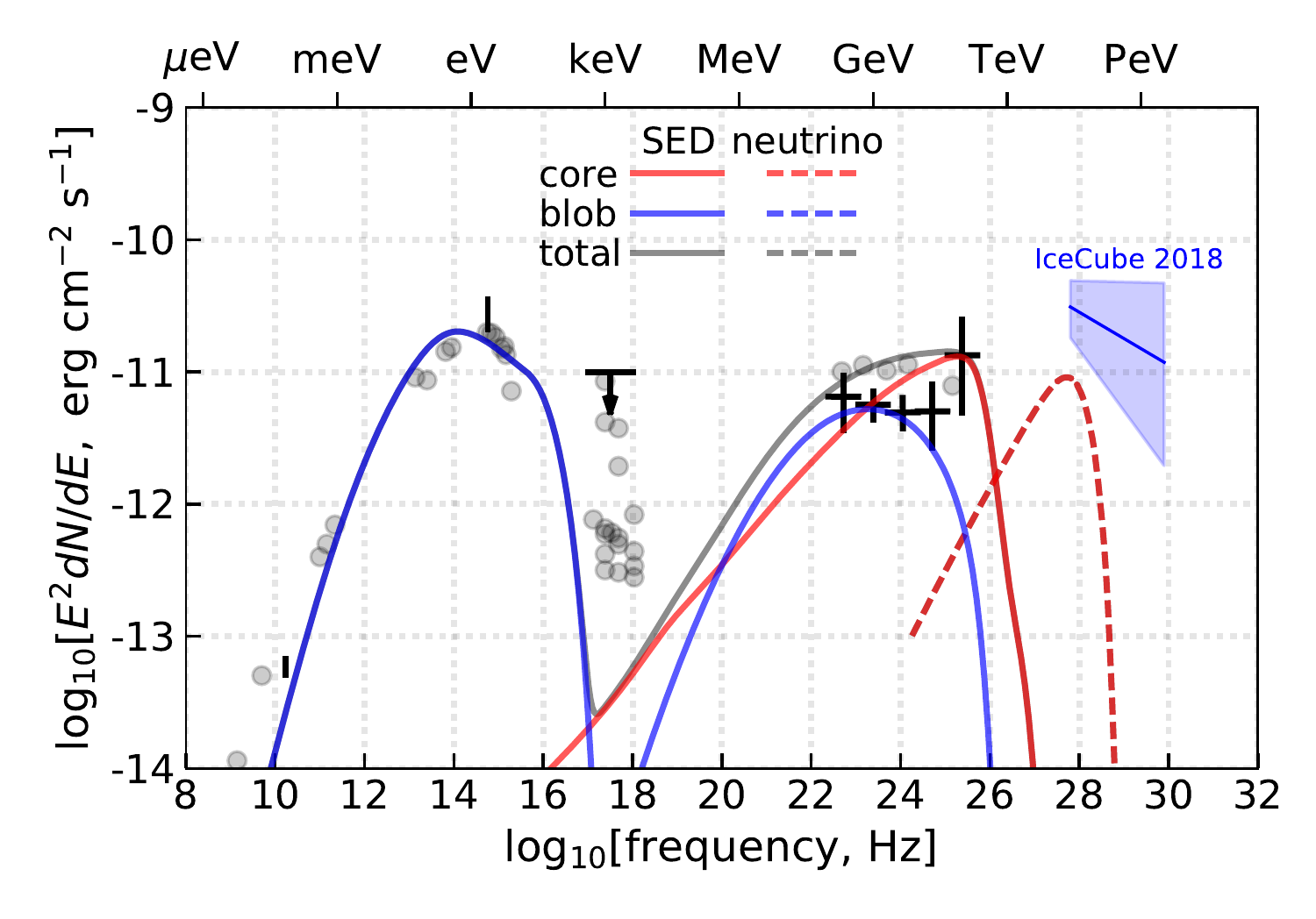}
	\caption{Emitted SED and muon neutrino spectrum from {\TXS{}} for one parameter set of the compact core model (\cf~\protect{\tabl{parameters}}). We plot the contributions from the blob region (blue), which accounts for the fluxes from radio to X-rays but does not produce neutrinos because it is of dominantly leptonic origin, and the contribution from the core region (red), which accounts for the neutrino flux and the gamma-ray data, separately. The parameter set was obtained through optimization of the emitted neutrino flux, which yields at most 1.9 IceCube events.}
	\label{fig:compactcore}
\end{figure}

As suggested earlier, a smaller emission region can enhance neutrino production. Following the model in~\citet{Gao:2018mnu} for the 2017 flare, we speculate that during the neutrino flare a compact core is formed inside the larger emission region, sharing its Doppler factor. This scenario can be regarded as a (spatially) structured jet model where the resulting emission of photons and neutrinos originates from the superposition of both radiation zones. In the present case, the core is a highly p$\gamma$-efficient region that simultaneously explains the gamma-ray and neutrino emission with a suppressed synchrotron cascade; most electromagnetic radiation at lower energies originates from the larger blob region. The so-called spine-sheath model assumes in addition a velocity structure that allows for finer control over the multi-messenger emission at the cost of a higher number of free parameters \citep{Ahnen:2018mvi}.

The results for the compact core model are shown in \figu{compactcore} for one optimized set of parameters; see also \tabl{parameters}. Emission from the blob describes the data from radio to soft gamma rays, while its contribution above GeV energies is low. The large volume of the blob translates into small target photon densities for photo-hadronic interactions, leading to inefficient neutrino production and a dim hadronic cascade (effectively leading to a leptonic model). The higher radiation densities in the core create an IC-dominated hadronic cascade, which accounts for a gamma-ray emission that hardens above 10 GeV. The suppression of synchrotron emission, in combination with small hadronic and Bethe-Heitler cascades, can suppress X-ray emission to a minimum. This example represents a neutrino-efficient model that is not strongly constrained by X-ray emission. Parameter sets can also be found that yield more neutrino events, but are in our view unphysical, \eg stronger magnetic fields in the blob than in the core. 

The compact core model {has also been previously applied to the 2017 flare \citep{Gao:2018mnu}.} However, it requires additional fine-tuning of the  core and the blob parameters to explain the temporal correlation among the optical, X-ray and gamma-ray flares. As a side effect, the compact core model can also reproduce a fast gamma-ray variability (for instance through a modulation of the core size that would have no effect on the radio to X-ray bands). However, the transition from a neutrino-quiescent to a neutrino-loud state without signature in gamma rays would require fine-tuning in the temporal evolution of the different parameters.
The neutrino flux emitted by the core translates to 1.9 observed muon tracks, which is slightly higher than in SED-compatible one-zone models; however, it is still in tension with the IceCube result.

\begin{figure}[btp]
	\centering
	\includegraphics[width=\columnwidth]{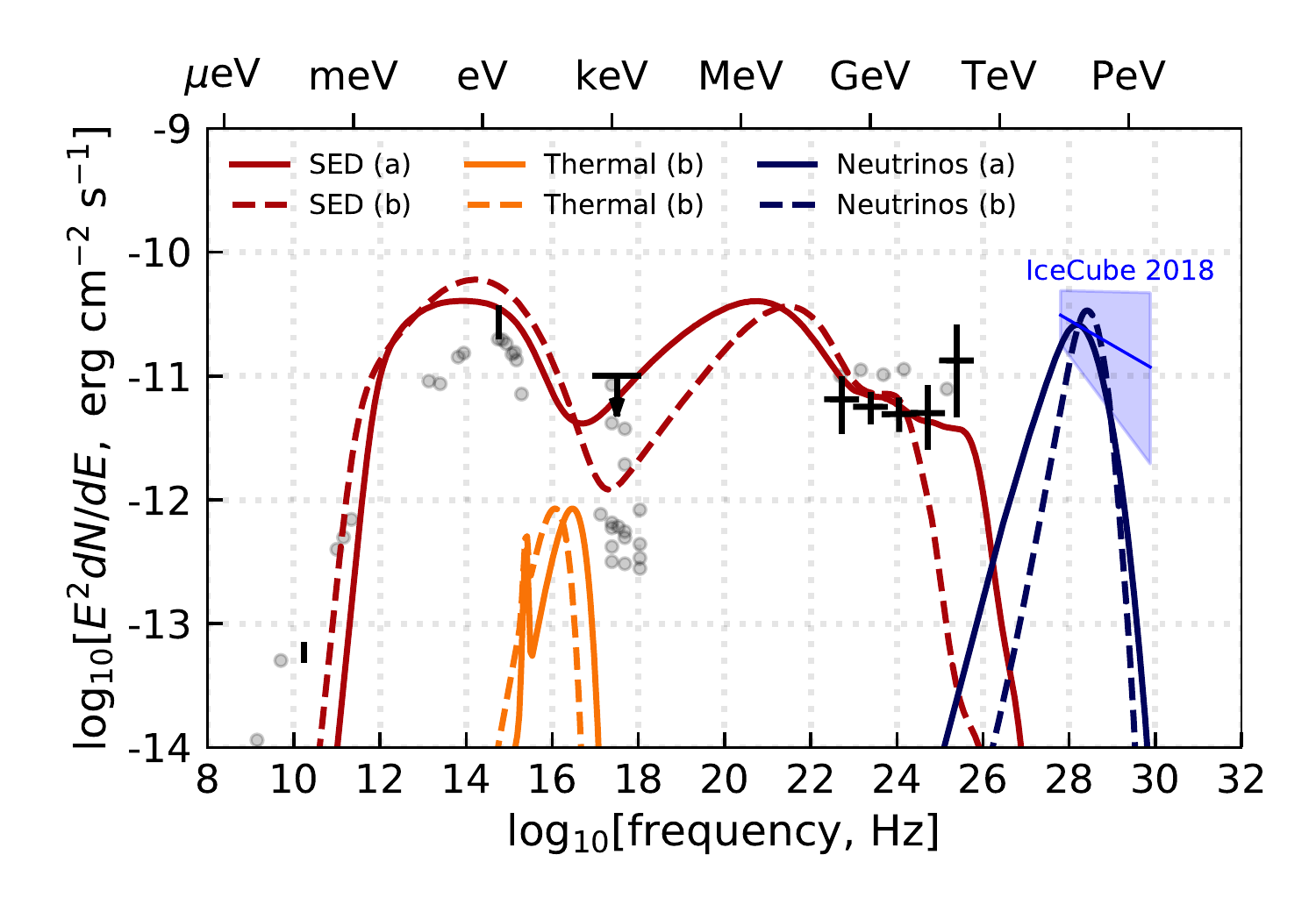}
	\caption{Emitted SED (red) and muon neutrino spectrum (blue) from {\TXS{}} when considering the contribution of external fields from the BLR, namely a thermal emission from the accretion disk and broad line emission (orange curves, shown in the observer's frame). The parameter sets (a) and (b), listed in \protect{\tabl{parameters}}, are shown as solid and dashed curves, respectively; their predicted number of muon tracks in IceCube is 4.9 and 4.0.}
	\label{fig:extfield}
\end{figure}

Finally, we {consider the impact} of an external thermal field, similarly to what has been assumed by~\citet{Keivani:2018rnh} {for the 2017 flare}. {While this source has been identified as a BL Lac due to the lack of significant broad line emission, \citet{Padovani:2019xcv} {argue that \TXS{} is a masquerading BL Lac that includes broad lines and thermal emission from an accretion disk as for FSRQs, which are outshined by the beamed non-thermal emission from the jet.}}

In that case, a fraction of the accretion disk radiation is isotropized through Thomson scattering in a BLR surrounding the disk (we fix this fraction to 1\%) {and a more significant fraction \citep[around 10\%,][]{Greene:2005nj} is re-emitted as atomic broad lines. If the {emitting region} lies within the BLR, these components} will appear boosted in the jet frame and interact with the non-thermal particles \citep[see \eg~][regarding the relativistic transformations]{Rodrigues:2017fmu}. {The thermal continuum is modeled as a blackbody emission of temperature $T_\mathrm{disk} \sim 10^5~\mathrm{K}$ \citep{Bonning:2006iq}; broad line emission is represented by the H$\alpha$ line, {which is} typically the brightest. The maximum proton energy is {adjusted such that interactions with the thermal continuum result in} neutrinos with energy $E_\nu\sim100~\mathrm{TeV}(T/10^5~\rm{K})^{-1}$}. Due to photon annihilation, the external fields also attenuate gamma rays from the jet as they cross the BLR, with maximum attenuation at $E_\gamma \sim 10~\mathrm{GeV}(E_\nu/100~\mathrm{TeV})^{-1}$. At this redshift, photons with energy $E>300$~GeV suffer additional loss due to the interaction with the extra-galactic background light (EBL, modeled in \citet{Dominguez11}), hence the steep cutoff of the SED shown in the figures.

The results of the external field model are presented in \figu{extfield}. The thermal field (orange) is out-shined by the highly beamed jet radiation (red) and is invisible to the observer. The solid red curve represents parameter set (a) in \tabl{parameters} and leads to {4.9} neutrino events during the flare. 
In this model, the hump observed in the optical band originates from synchrotron emission by $e^\pm$ pairs from Bethe-Heitler production, while the hump in the MeV-GeV range is emitted by $e^\pm$ pairs from the annihilation of hadronic photons. The emission from primary electrons is therefore sub-dominant across the spectrum, which implies that it is difficult to identify such a model in simplified (such as analytical) approaches. However, as mentioned above, the high neutrino production efficiency implies a softening and a suppression of the gamma-ray spectrum above 10~GeV. {A} spectral softening is in tension with \textit{Fermi} observations \citep{Aartsen:2019gxs,Padovani:2018acg}. The X-ray bound is almost saturated, as well. One of the specifics of this model is the anti-correlation between VHE gamma-ray and neutrino emission, which has been previously discussed in \citet{Murase:2015xka} (see also \citet{2018arXiv181202398X}). The parameter {set (b) in \figu{extfield} yields 4 neutrino events; it has a slightly lower X-ray component at the cost of a higher tension with the last \textit{Fermi} data point.} The attenuation of high-energy gamma rays between models depends not only on the disk luminosity but also on the assumed radius of the BLR (\cf~\tabl{parameters}). \revise{Note that given the disk luminosity of the source, phenomenological relationships would suggest a BLR radius of around $3\times10^{17}~\mathrm{cm}$ \citep{Kaspi:1999pz}. In the examples shown, the values of $R_\mathrm{BLR}$ differ from that reference value by a factor of two or less, which is within the statistical spread of the AGN sample reported by \citet{Kaspi:1999pz}.}

\section{Summary and conclusions}

We have tested the compatibility of the historical 2014-15 neutrino flare from the blazar \TXS{} with leptohadronic (photohadronic) multi-messenger source models. Within the constraints of the sparse observations in optical, X-rays and gamma rays during the neutrino flare, we scanned the parameter space using several distinct assumptions about the geometry and environment of the blazar. In addition to conventional one-zone models,
we have considered scenarios involving a compact core emission region (corresponding to a spatially structured jet) and external radiation fields, such as that from an accretion disk. 

We have demonstrated that at most {two to five} neutrino events during the period of the flare can be expected from any of the three models {in compatibility} with multi-wavelength constraints. While the one-zone model saturates the available X-ray bound, the SED at gamma-ray energies can be reasonably reproduced. The electromagnetic cascade from charged and neutral pion decays can be hidden as a prominent hump at MeV energies, where no data is available. 

A compact core model yields a similar expectation for the neutrino rate as the most optimistic one-zone model, accompanied by a spectral hardening in gamma rays above 10~GeV. The radiation at highest energies and the neutrinos both originate from the inverse-Compton-dominated core, whereas the X-ray data are generated by a larger emission region via a leptonic synchrotron self-Compton (SSC) process. A natural feature is a faster gamma-ray variability from the compact core compared to the slower radio-to-X-ray variability. However, a transition between neutrino-quiescent and flaring states would imply a fine-tuned correlation in the evolution of the parameters.

The external radiation field model {yields SEDs with more than two neutrino events} during the flare; however, the high neutrino production efficiency implies a higher optical thickness to $\gamma\gamma$ annihilation at VHEs, softening the expected gamma-ray spectrum -- in {minor} tension with \textit{Fermi} observations. Such a softening or cutoff can serve as an electromagnetic signature for an orphan neutrino flare.

While we do not claim completeness, our study demonstrates the obstacles involved in the simultaneous description of the electromagnetic SED and neutrino observations during the historical flare. Since all present models are challenged by the high neutrino event rate, we have not even attempted to describe the temporal evolution, \ie{} the transition between the neutrino-quiescent and flaring states, or to achieve a unified description between the 2014-15 and 2017 flares. However, from our modeling and extensive parameter scans, we conclude that a) obtaining more than two to five neutrino events during the flare implies violating multi-wavelength constraints, particularly in gamma rays, and that b) a transition between the neutrino-quiescent and flaring states without distinctive electromagnetic activity is unlikely within the photohadronic framework. Let us finally remark that the $13 \pm 5$ signal events quoted by IceCube are obtained under the assumption of a power-law spectrum with a spectral index $-2.2$. For a harder spectrum or even a different spectral shape, the signal emerging over the atmospheric background may be {significantly} smaller.

\acknowledgments

The authors would like to thank an anonymous reviewer for their useful comments. This work has been supported by the European Research Council (ERC) under the European Union’s Horizon 2020 research and innovation programme (Grant No. 646623).

\bibliographystyle{aasjournal}

\end{document}